\documentclass[
twocolumn,
]{ceurart}

\usepackage{listings}

\def\eqref#1{equation~\ref{#1}}

\def\1{\boldsymbol{1}}

\def\rmX{{\mathbf{X}}}

\def\ermX{{\textnormal{X}}}

\def\va{{\boldsymbol{a}}}
\def\vb{{\boldsymbol{b}}}

\def\eva{{a}}
\def\evb{{b}}

\def\mC{{\boldsymbol{C}}}

\def\mP{{\boldsymbol{P}}}

\def\mS{{\boldsymbol{S}}}

\def\mU{{\boldsymbol{U}}}

\DeclareMathAlphabet{\mathsfit}{\encodingdefault}{\sfdefault}{m}{sl}
\SetMathAlphabet{\mathsfit}{bold}{\encodingdefault}{\sfdefault}{bx}{n}

\def\sN{{\mathbb{N}}}

\def\emC{{C}}

\def\emP{{P}}

\def\emS{{S}}

\def\emU{{U}}

\usepackage{amssymb,amsmath,amsthm}
\usepackage{mathtools} %
\usepackage{subcaption}
\usepackage{caption}
\usepackage{soul}
\usepackage{multirow}
\usepackage{float}
\usepackage{enumitem}

\makeatletter
\DeclareRobustCommand{\iscircle}{\mathord{\mathpalette\is@circle\relax}}
\newcommand\is@circle[2]{%
  \begingroup
  \sbox\z@{\raisebox{\depth}{$\m@th#1\bigcirc$}}%
  \sbox\tw@{$#1\square$}%
  \resizebox{!}{\ht\tw@}{\usebox{\z@}}%
  \endgroup
}
\makeatother

\makeatletter
\newcommand*{\inlineequation}[2][]{%
  \begingroup
    \refstepcounter{equation}%
    \ifx\\#1\\%
    \else
      \label{#1}%
    \fi
    \relpenalty=10000 %
    \binoppenalty=10000 %
    \ensuremath{%
      #2%
    }%
    ~\@eqnnum
  \endgroup
}
\makeatother

  \newtheorem{theorem}{Theorem}[section]

  \newtheorem{proposition}[theorem]{Proposition}

  \newtheorem{definition}[theorem]{Definition}

\begin{document}

\copyrightyear{2022}
\copyrightclause{Copyright for this paper by its authors.
  Use permitted under Creative Commons License Attribution 4.0
  International (CC BY 4.0).}

\conference{RecSys in HR'23: The 3rd Workshop on Recommender Systems for Human Resources, in conjunction with the 17th ACM Conference on Recommender Systems, September 18--22, 2023, Singapore, Singapore.}

\title{FEIR: Quantifying and Reducing Envy and Inferiority for Fair Recommendation of Limited Resources}

\author[1]{Nan Li}[%
email=nan.li@ugent.be,
]

\author[1]{Bo Kang}[%
email=bo.kang@ugent.be,
]

\author[1]{Jefrey Lijffijt}[%
email=jefrey.lijffijt@ugent.be,
]

\author[1]{Tijl {De Bie}}[%
email=tijl.debie@ugent.be,
]
\address[1]{Ghent University, Ghent 9000, Belgium}

\cortext[1]{Corresponding author.}

\begin{abstract}
In settings such as e-recruitment and online dating, recommendation involves distributing limited opportunities, calling for novel approaches to quantify and enforce fairness. We introduce \emph{inferiority}, a novel (un)fairness measure quantifying a user's competitive disadvantage for their recommended items. Inferiority complements \emph{envy}, a fairness notion measuring preference for others' recommendations. We combine inferiority and envy with \emph{utility}, an accuracy-related measure of aggregated relevancy scores.
Since these measures are non-differentiable, we reformulate them using a probabilistic interpretation of recommender systems, yielding differentiable versions. We combine these loss functions in a multi-objective optimization problem called \texttt{FEIR} (Fairness through Envy and Inferiority Reduction), applied as post-processing for standard recommender systems.
Experiments on synthetic and real-world data demonstrate that our approach improves trade-offs between inferiority, envy, and utility compared to naive recommendations and the baseline methods.
\end{abstract}

\maketitle

\section{Introduction}\label{intro}
Fairness in machine learning based recommendation systems attracts increasing research attention, driven both by ethical and legal motivations.
Here we focus on recommending items with \textbf{limited availability}, such as job recommendation, online dating, and education resource recommendation. The need for users to \emph{compete} for recommended items distinguishes this recommendation setting from more standard ones such as e-commerce, or movie or music recommendation, where items have practically unlimited availability.

When multiple users are recommended the same item, they enter a competition for that item. Only one or a few will win and obtain the item, leaving the others empty-handed. For example, a job seeker who applies for their recommended jobs could fail to get employed if these jobs were also recommended to better qualified rivals. 
This competition aspect brings specific challenges to evaluate and improve the fairness of recommendation strategies---challenges that have hitherto not been recognized.

To discuss this setting, it is useful to consider two possibly distinct kinds of affinity between a user and an item: an item's \textbf{utility} for the user (i.e. the user's preference), and a user's \textbf{suitability} (i.e. competitiveness) for the item. In traditional recommender systems, only utility is relevant, as suitability is directly related to the competitive nature of the setting.

We consider two ways in which unfairness can arise in such settings.
First, a fair recommendation system for limited resources should ensure that users prefer their own recommended items over those recommended to others. 
This idea is captured by the notion of \textbf{envy}: user A has \emph{envy} towards user B if the utility of user B's recommendations in user A's perspective is higher than user A's recommendations.
Second, it is also arguably unfair to an individual if her recommended items are \emph{always} also recommended to people more suitable to them than her, as she would fruitlessly compete for it.
This idea is captured by the notion of \textbf{inferiority}: user A is \emph{inferior} to user B if A is less suitable than B to the items recommended to both A and B. We argue that a fair recommender system in this setting should yield low envy and low inferiority for everyone.

As an illustration, consider the scenario of two users, \textbf{1} and \textbf{2}, and three items, $\iscircle$, $\square$, and $\triangle$. Let the \emph{utility} scores be represented by the matrix: $\begin{bmatrix}  
\iscircle 0.2 & \square 0.6 & \triangle 0.9 \\[-3pt]
\iscircle 0.1 & \square 0.8 & \triangle 0.7 
\end{bmatrix}$, where the first row represents user \textbf{1}'s scores and the second row represents \textbf{2}'s scores. Similarly, let the \emph{suitability} scores, or the chances of a user getting the item, be represented by the matrix: $\begin{bmatrix}
\iscircle 0.3 & \square 0.9 & \triangle 0.4 \\[-3pt]
\iscircle 0.3 & \square 0.8 & \triangle 0.8 
\end{bmatrix}$.

Examples: Recommending only $\triangle$ to both users results in no envy, as the recommendations are equivalent and thus neither user prefers the other's. However, there is high inferiority, as \textbf{1} is less suitable than \textbf{2}, and thus less likely to obtain $\triangle$. Recommending $\iscircle$ to \textbf{1} and $\square$ to \textbf{2} results in high envy as \textbf{2}'s recommendation has higher utility for \textbf{1} than their own recommendation, but no inferiority, as both users are recommended an item that is only recommended to themselves. Recommending $\iscircle$ to both users results in neither envy, nor inferiority, but has low utility for both users. What is the best recommendation in this case depends on the chosen trade-off.

The illustration above shows that both fairness notions are necessary: minimizing inferiority tends to result in less preferred jobs being recommended, which, if left uncontrolled, risks increasing envy. Moreover, there is also a trade-off between utility and both notions of fairness, particularly but not exclusively with inferiority.

Given the high stakes involved in many applications of this setting (with job recommendation as a notable example), there is an urgent need to adopt these notions of fairness in practical applications. While there is some work on the related notion of \emph{congestion} and some limited work has been done on envy in recommender systems (see Sec.~\ref{sec:related_work}), we are unaware of any research directly addressing this need. This paper fills that gap, by formalizing these concepts as well as by proposing the \texttt{FEIR} (\textbf{F}airness through \textbf{E}nvy and \textbf{I}nferiority \textbf{R}eduction) method for post-processing the results of any other recommendation algorithm, yielding recommendations with low envy and low inferiority, while still maintaining high utility. Our specific contributions are:
\begin{enumerate}
\item We propose and formalize inferiority as a new individual fairness concept that is complementary to envy, when recommending items with limited availability. To facilitate minimizing these notions, we also derive their expected values with respect to a probabilistic interpretation of recommendation algorithms, resulting in differentiable versions. (Sec.~\ref{sec:quantification}.)
\item Leveraging these differentiable versions, we propose the \texttt{FEIR} algorithm, a model-agnostic post-processing method of the output scores of any upstream recommendation algorithm for all user-item pairs. \texttt{FEIR} seeks a fairer score matrix by solving a multi-objective optimization problem with the goal of minimizing the expected envy and inferiority, and maximizing the expected utility. (Sec.~\ref{sec:optimization}.)
\item We investigate \texttt{FEIR}'s ability to trade-off both fairness measures and utility in extensive experiments both on synthetic and real data. We also demonstrate superiority of \texttt{FEIR} compared with the baseline methods. (Sec.~\ref{sec:experiments}.) 
\end{enumerate}

\section{Method}\label{sec:method} 
In this section, we first give quantifications of utility, envy and inferiority in the \emph{deterministic} setting and the \emph{probabilistic} setting (Sec.~\ref{sec:quantification}). Second, we formulate the problem of finding a good recommendation strategy as a multi-objective optimization problem solvable by minimizing a weighted sum of loss terms, leading to the \texttt{FEIR} method (Sec.~\ref{sec:optimization}).

\subsection{Quantification}\label{sec:quantification}
Let $\va = (\eva_1, \dots, \eva_m)$ be $m$ users, $\vb = (\evb_1, \dots, \evb_n)$ be $n$ items. A recommender system recommends $k$ items to every user.
The utility matrix $\mU$ is an $m \times n$ matrix, where each entry $\emU_{i,j} \in (0,1)$ represents the utility of item $\evb_j$ to user $\eva_i$, so that each row $\mU_{i,:}$ represents the utility function of $\eva_i$ evaluated on all the items. The suitability matrix $\mS$ is also an $m \times n$ matrix where each entry $\emS_{i,j} \in (0,1)$ represents the suitability (matching degree), between user $\eva_i$ and item $\evb_j$. 
 
\subsubsection{Deterministic setting}\label{sec:deterministic_setting}
$\mU$ gives us the item-wise utility, but in recommendation we need to measure the utility of a list of $k$ items. Note that this list is a $k$-sized multiset constructed from $\vb$ with \emph{repeated recommendations allowed}, although in practice the chance of repetition is very slim when $k \ll n$. Another motivation of allowing repetition is for mathematical convenience as shown in \ref{def:expected_u_e_i}.

Let $\mC^k$ be an $m \times n$ counting matrix where each entry $\emC^k_{i,j} \in \sN^0$ is the number of occurrences of job $\evb_j$ in the recommendation for $\eva_i$. Then each row $\mC^k_{i,:}$ represents the recommendation list for $\eva_i$ such that $\sum_{j=1}^n \emC^k_{i,j} = k$ for all $i$. We omit the superscript $k$ if the context is clear.

\begin{definition}[User utility]\label{def:discrete_u_individual}
The \textbf{utility} of the recommendation for job seeker $\eva_i$ is a simple summation of the utility of each job to $\eva_i$ in $\eva_i$'s list: 
$$u(\eva_i, \mU, \mC_{i,:}) = \sum_{j=1}^n \emU_{i,j}\emC_{i,j}.$$
\end{definition}

Envy measures the comparative utility from each individual's perspective. It captures the idea that an individual may feel envy towards another if another person's recommended items have higher utility to them, wrt. their own utility.\footnote{If repetition is not allowed, the formulation would use an indicator function representing whether item $j$ is recommended to user $i$. The major drawback of this formulation is that its probabilistic counterpart (the process of sampling without replacement) follows the hypergeometric distribution, which is computationally difficult. Further investigation is left future work.}

\begin{definition}[User envy]\label{def:discrete_e_individual}
The \textbf{envy} from $\eva_i$ to $\eva_{i^*}$ is:
$$e(\eva_i,\eva_{i^*},\mU,\mC) = \sum_{j=1}^n \emU_{i,j}(\emC_{i^*, j} - \emC_{i,j}).$$
\end{definition}

Inferiority represents the disadvantage of one user to another when they compete for the same items, such as applying for the same jobs. It is measured based on the suitability between users and items, represented by the matrix $\mS$.

\begin{definition}[User inferiority]\label{def:discrete_i_individual}
The \textbf{inferiority} from $\eva_i$ to $\eva_{i^*}$ is:
\begin{align*}
f(\eva_i,\eva_{i^*},\mS, \mC) &= \sum_{j=1}^n \max(0, \emS_{i^*,j} - \emS_{i, j}) \\
&\quad \cdot\min(1, \emC_{i, j}\emC_{i^*, j}).
\end{align*}
\end{definition}
Inferiority captures the difference in suitability between user $\eva_i$ and $\eva_{i^*}$ towards all the \emph{common} recommended items, i.e., inferiority is only concerned with items recommended to both users. Note that, if any item occurred more than once, we only count it once, this is to consider the competition between them over the same item only once.

\begin{definition}\label{def:discontinousmeasures}
Utility, envy, and inferiority on the \emph{system} level are simply the averages of the positive user-level measurements:
\begin{align}
u(\va, \mU, \mC) &=  \tfrac{1}{m} \sum_{i=1}^m u(\eva_i, \mU, \mC),\label{eq:discrete_u_system}\\
e(\va, \mU, \mC) &= \tfrac{1}{m} \sum_{1 \le i \ne i^* \le m } \max\left(0, e\left(\eva_i,\eva_{i^*},\mU,\mC\right)\right), \label{eq:discrete_e_system}\\
f(\va, \mS, \mC) &= \tfrac{1}{m} \sum_{1 \le i \ne i^* \le m }  f(\eva_i,\eva_{i^*},\mS,\mC) \label{eq:discrete_i_system}.
\end{align}
\end{definition}

The $\max(0,\cdot)$ in the definition of the envy ensures that only positive contributions are counted, to avoid a negative envy in one user to compensate a positive envy in another. (Individual user utilities and inferiorities are always positive.)

\subsubsection{Probabilistic setting}\label{sec:probabilistic_setting}
The discontinous nature of recommender systems as recommending multisets of items makes it practically impossible to utilize the utility, envy, and inferiority from Def.~\ref{def:discontinousmeasures} in an optimization-based approach. We will thus develop probabilistic alternatives that are differentiable.

We consider a probabilistic recommendation setting in which the recommendation strategy is represented by a user-item mapping function, $\pi: \va \times \vb \rightarrow [0,1]$, that assigns a probability to each user-item pair of the recommendation of the item to the user. To model the recommendations, we assume an independent multinomial process for each user, where each user has a different $n$-sided uneven dice, and to recommend $k$ ($k \ll n$) items, we throw the dice $k$ times and take the outcome as the recommendation.

Then a recommender strategy can be represented as an $m \times n$ matrix $\mP\in[0,1]^{m\times n}$ where each entry $\emP_{i,j}$ is the probability of recommending $\evb_j$ to $\eva_i$.
All users' $k$-sized recommendation can be written as a random matrix $\rmX$ where each row $\rmX_{i,:}$ is a random vector for $\eva_i$ where the random variables $\ermX_{i,j}$ indicate the number of times item $\evb_j$ is included in $\eva_i$'s list. By our setting $\rmX_{i,:}$ follows a multinomial distribution with parameters $k$ and $\mP_{i,:}$.

In this context of probabilistic recommendation, the \emph{expected values} of a user's utility, envy, and inferiority are given by the following Proposition:
\begin{proposition}[Expected user utility, envy, and inferiority]\label{def:expected_u_e_i}
\begin{align}
&\mathbb{E}_{\rmX \sim \mP} [u(\eva_i, \mU, \rmX_{i,:})]  = k \sum_{j=1}^n \emP_{i,j} \emU_{i,j},\label{eq:stochastic_u_individual}\\
&\mathbb{E}_{\rmX \sim \mP} [e(\eva_i, \eva_{i^*}, \mU, \rmX)] = k \sum_{j=1}^n (\emP_{i^*, j} - \emP_{i,j})\emU_{i,j},
\label{eq:stochastic_e_individual}\\
&\mathbb{E}_{\rmX \sim \mP}[f(\eva_i, \eva_{i^*},\mS, \rmX)] \nonumber \\
&\quad=  \sum_{j=1}^n \max(0, \emS_{i^*, j} - \emS_{i, j} ) \nonumber \\
&\quad\quad \cdot(1-(1-\emP_{i, j})^k)(1-(1-\emP_{i^*, j})^k). \label{eq:stochastic_i_individual}
\end{align}
\end{proposition}

\begin{proof}[Proof outline]
For utility and envy, this follows from the fact that $\mathbb{E}_{\rmX \sim \mP}[X_{i.j}]=kP_{i,j}$ (the factor $k$ stemming from $\sum_{i=1}^n x_i = k$), and from linearity of the expectation operator.
For inferiority, this follows from linearity of the expectation operator, and from the fact that $(1-(1-\emP_{i, j})^k)(1-(1-\emP_{i^*, j})^k)$ is the probability that both $C_{i,j}$ and $C_{i*,j}$ are non-zero integers, and thus the probability that $\min(1,C_{i,j}C_{i*,j})$ is equal to 1.
\end{proof}

For a recommendation system for users $\va$  over items $\vb$, represented by the $m \times n$ matrix $\mP$, the expected utility, envy and inferiority of the $k$-sized recommendation $\rmX$ on the \emph{system} level is the average of the expected values of all users:
\begin{align}
&\mathbb{E}_{\rmX \sim \mP}[u(\va, \mU, \rmX)] =
\tfrac{1}{m}\textstyle\sum_{i=1}^m \mathbb{E}_{\rmX \sim \mP} [u(\eva_i, \mU, \rmX_{i,:})],\label{eq:expected_utility}\\
&\mathbb{E}_{\rmX \sim \mP}[e(\va, \mU, \rmX)] = \tfrac{1}{m}\textstyle\sum_{1 \le i \ne i^* \le m } 
\max(0,\nonumber \\
&\qquad\qquad\qquad\qquad\qquad\mathbb{E}_{\rmX \sim \mP}[e(\eva_i, \eva_{i^*}, \mU, \rmX)]),\label{eq:expected_envy}\\
&\mathbb{E}_{\rmX \sim \mP}[f(\va, \mS, \rmX)] = \tfrac{1}{m}\textstyle\sum_{1 \le i \ne i^* \le m } \nonumber \\ &\qquad\qquad\qquad\qquad\qquad\mathbb{E}_{\rmX \sim \mP} [f(\eva_i, \eva_{i^*}, \mS, \rmX)] \label{eq:expected_inferiority}.
\end{align}

\subsection{Optimization by minimizing combined losses}\label{sec:optimization}
The \texttt{FEIR} algorithm minimizes a combined loss function defined from the expected utility, inferiority, and envy of the recommendation system. It uses a gradient descent based method to optimize the scores of the resulting recommendation strategy, represented by the matrix $\mP'$, by solving:
\begin{equation}\label{eq:total_loss}
\begin{aligned}
 \ell_{total}(\mP', \mS, \mU) &= w_1 \ell_{e}(\mP', \mU) \\
 &+ w_2 \ell_{f}(\mP', \mS) \\
 &+ w_3 \ell_{u}(\mP', \mU) \\
 &+ w_4 \ell_{p}(\mP'),
\end{aligned}
\end{equation}
where $\ell_{e}(\mP', \mU)$, $\ell_{f}(\mP', \mS)$, $\ell_{u}(\mP', \mU)$, and $\ell_{p}(\mP') = \sum_{i=1}^m (\sum_{j=1}^n \emP'_{i,j} -1 )^2$ are the expected envy (Eq.~\ref{eq:expected_envy}), expected inferiority (Eq.~\ref{eq:expected_inferiority}), negative expected utility (Eq.~\ref{eq:expected_utility}), and a penalty term for making each row of $\mP'$ a probability distribution, respectively. Parameters $w_1$, $w_2$, $w_3$, $w_4$ are weights for each term. The penalty term $\ell_{p}(\mP')$ can be omitted if the matrix $\mP'$ is renormalized after each update (e.g., using row-wise softmax as the activation function). An important benefit of \texttt{FEIR} is its being model-agnostic: any model capable of scoring all user-item pairs can be post-processed by \texttt{FEIR}.

\emph{Notes on probabilistic and deterministic settings.}  
The probabilistic setting is more general and mathematically convenient, but real recommendation systems typically recommend (deterministically) each user the $k$ items with the highest probabilities. Thus, our experiments train with the probabilistic but evaluate with the deterministic measures.

\emph{Notes on available affinity types in current systems.} 
In our definitions, $\mS$ represents the suitability of users to items, and $\mU$ quantifies the utility of items to users. The difference between them can create tension between envy and inferiority, as seen when a job seeker prefers unsuitable jobs. 
However, real-world applications typically use a single affinity score provided by an existing recommender system, combining both suitability and utility. 
Thus, for practical reasons, in our experiments only one set of affinities is used to calculate both envy and inferiority, except for one synthetic dataset. 
In these cases, utilities and suitabilities align, but tension between envy and inferiority still arises due to individual differences in scores. This is illustrated by a toy example where the scores for users \textbf{1} and \textbf{2} with respect to items are $\begin{bmatrix}
\iscircle 0.1 & \square 0.9 & \triangle 0.8 \\[-3pt]
\iscircle 0.4 & \square 0.6 & \triangle 0.5 
\end{bmatrix}$, and recommending $\square$ to both of them results in no envy and high utility, but high inferiority from \textbf{2} to \textbf{1}.

\subsection{Scaling-up methods}
\label{sec:scaling-up}
With large scale data, we propose the following approximation methods: inferiority loss mini-batching, user sampling, item sampling, user-item sampling.
\emph{Mini-batching} randomly splits the $m$ users into $\lfloor m/b \rfloor$ batches and at each step calculates the inferiority from $b$ users within the current mini-batch to all the users with respect to all the items, leaving other losses calculated globally. \emph{User sampling} takes a random subset of $m_s$ users at each training step and calculates the losses within this subset. \emph{Item sampling} takes a random subset of $n_s$ items at each training step and calculates the losses between all user pairs with respect to only those items. \emph{User-item sampling} samples from both the users and items at each training step. 

\subsection{Metrics}\label{sec:metrics}
We evaluate recommendation strategies based on the one-time deterministic recommendation obtained from the probabilistic strategy. For different $k$s, let $\mC^k$ be the binary matrix obtained from the recommendation strategy $\mP$ by setting the item indices with the highest $k$ values in each row to be 1 and the rest to be 0. 

\emph{Normalized system-level utility and fairness.} The system-level utility, envy and inferiority for top-$k$ recommendation are defined by Eq.~\ref{eq:discrete_u_system}, \ref{eq:discrete_e_system} and \ref{eq:discrete_i_system}, albeit $\mU = \mS$ in the experimental data. We also calculate the \textbf{overall fairness} as 
$$g(\va, \mU, \mS, \mC^k) = e(\va, \mU, \mC^k) + f(\va, \mS, \mC^k).$$ 
Let $C_{naive}^K$ denote the naive recommendation, then we have the \textbf{normalized system-level top-$k$ recommendation metrics} defined as 
$$\frac{\psi(\va, \mU, \mC^k)}{\psi(\va, \mU, \mC_{naive}^k)},$$
where $\psi \in \{u, f, g\}$ (no normalized envy since $e(\va, \mU, \mC_{naive}^k)=0$).

\emph{Competition faced by each user.} To address \ref{enum:rq2}, we use the following competition indicators. 
The \textbf{mean rank} of job seeker $\eva_i$ is calculated as: 
	$$\text{rank}(i) := \frac{1}{k} \sum_{j=1}^n |D_{i,j}|,$$
	where $$D_{i,j} = \{\eva_{i^*} | \emC_{i,j} = \emC_{i^*,j}=1, \emS_{i^*,j} > \emS_{i, j} \},$$ which measures the average rank of user $\eva_i$ among her competitors for the same recommended items.
The \textbf{mean suitability gap} of user $\eva_i$ is calculated as: 
	$$\text{gap}(i) = \frac{1}{k} \sum_{j=1}^n \emC_{i,j} \frac{1}{\max(1, |D_{i,j}|)}\sum_{i^* \in D_{i,j}}(\emS_{i^*, j} - \emS_{i, j}),$$ which measures the average difference in suitability scores between user $\eva_i$ and her likelier competitors for the same recommended items. 
By averaging these metrics over all users, we can obtain an overall evaluation of the competition.

\emph{Multiple solutions comparison.} For methods that can generate multiple solutions representing different levels of trade-offs, we plot the Pareto frontiers to visually compare sets of solutions. Additionally, we use the following numerical metrics:
\begin{enumerate}
\item \textbf{HV} (hypervolume) measures the amount of the objective space (relative to a reference point) that is dominated by the points on the frontier.
\item \textbf{Fairness above utility threshold} $\min(\phi|t)$: the minimum value of a fairness metric $\phi$ among all solutions with utility higher than $t$, where $\phi$ could be inferiority, overall fairness, mean rank or mean suitability gap. This allows us to assess how well a solution performs in terms of fairness for a given level of utility. %
\end{enumerate}

\emph{Item-side fairness.} Although our focus is user-side fairness, we also investigated the item-side fairness by comparing the Gini index of item exposure (\cite{mansoury2020fairmatch, mansoury2021graph, ge2021longterm,do2022optimizing}) before and after \texttt{FEIR} post-processing. 

\section{Experiments}\label{sec:experiments}
To evaluate the effectiveness of \texttt{FEIR}, we conduct experiments to answer the following research questions:
\begin{enumerate}[label=\textbf{RQ\arabic{enumi}}.,ref=RQ\arabic{enumi}]
\item \label{enum:rq1} How does \texttt{FEIR} compare to the baseline methods in improving the \emph{trade-offs between envy, inferiority and utility}?
\item \label{enum:rq2} Does \texttt{FEIR} decrease the \emph{competition measurements defined from rivals} compared to the baseline methods?
\end{enumerate}

\subsection{Datasets}\label{sec:dataset}
In our experiments, we use a variety of synthetic and real-world datasets to evaluate the performance of our proposed method. There are three types of synthetic datasets.
\textbf{Random synthetic data with distinct suitability and utility (SU50)}: Two $50 \times 50$ real-numbered matrices generated from a truncated normal distribution $(0,1)$ representing suitability scores and utility scores for 50 users and 50 items.  
\textbf{Random synthetic data with one set of scores}: Matrices generated from a truncated normal distribution $(0,1)$ with \emph{varying ratios} of number of users and items to investigate the effect of varying these ratios.
\textbf{Structured synthetic data with one set of scores}: Two $20 \times 100$ real-numbered matrices that simulate specific scenarios: Item groups (IG) and User groups (UG). The \textbf{IG} dataset represents the scenario where certain items have generally higher scores across all users, while the \textbf{UG} dataset represents the scenario where certain users have generally higher scores across all items.

We also use four real-world datasets, all obtained from the same upstream job recommendation model based on \cite{kang2018conditional}. 
\textbf{Zhilian}: Scores for 2,781 users and 6,568 items, sampled from a public dataset provided by a Chinese online recruitment platform.
\textbf{CareerBuilder}: Scores for 7,459 users and 11,020 items, obtained from a public dataset provided by \textit{CareerBuilder}.
\textbf{VDAB small}: Scores for 1,186 users and 8,921 items, a random sample from a private dataset provided by a labor agency in Belgium.
\textbf{VDAB large}: Scores for 10,369 users and 66,898 items, a random sample from the same source as VDAB small, but including more data. Results of this dataset are omitted due to space limitation.\footnote{Due to the size of \textbf{VDAB large}, we experimented several methods for scaling up, including sampling and mini-batching. The results are included in our online supplementary materials.}

\subsection{Baselines}\label{sec:baseline}
We use the following four baseline methods.

\emph{Standard recommendation (Naive).} Given the scores between all users and items, the common practice is to recommend the items with the highest $k$ scores to each user. 

\emph{Randomization of top scored items (Shuffle).} Randomly Sample $k$ items from items with the top-$d$ ($\ge k$) scores.

\emph{Congestion alleviation method (CA).} \citet{naya2021designing} proposes a congestion alleviation method based on linear optimization that aims to decreases the competition in the job market by using optimal transport. 
CA casts the problem of minimizing congestion into a linear program where the objective is to maximize the element-wise product of the original probability matrix and the solution matrix under the constraint of evenly distributing the probability of recommending each item. 

\emph{Modified Round-Robin procedure (RR).} Modified based on \cite{patro2020fairrec}, RR sets a threshold $\tau$ for suitability, randomly orders the users and then in each round, allocates one item for each user at each round such that this item is the most preferred one for this user with suitability greater than $\tau$. $k$ rounds would be run for top-$k$ recommendation. Unlike the other methods, RR is applicable only when $U$ and $S$ are both available.

\subsection{Experiment setting}\label{sec:experiment_setting}
For our method \texttt{FEIR}, we initialize the parameters by applying a row-wise softmax to the given scores and use gradient descent based methods to minimize the loss function defined in Eq.~\ref{eq:total_loss}. We perform a coarse search to find an appropriate learning rate, and then use this value to train the model with different combinations of loss weights to achieve different trade-offs between envy, inferiority and utility.
For the CA baseline, different entropic relaxation terms are used to roughly controls the trade-offs.
For the synthetic datasets, we train and evaluate strategies for the top 10 recommendation. For the real-world datasets, we train and evaluate strategies for different $k$s, ranging from 1 to 100. For the VDAB large dataset only a medium size $k=20$ is trained and evaluated due to time limitations.

We explore all scaling-up methods with the VDAB small dataset with $k=100$, find all methods perform similarly besides item sampling. Therefore, we apply one method to each real-world dataset for a full range of $k$s: mini-batching to the VDAB small dataset, user sampling to the Zhilian and Careerbuilder datasets, user-item sampling to VDAB large.

\begin{figure*}[t]
    \centering
    \captionsetup{font=footnotesize}
    \begin{subfigure}[t]{0.24\textwidth}  
        \centering 
        \includegraphics[height=26mm]{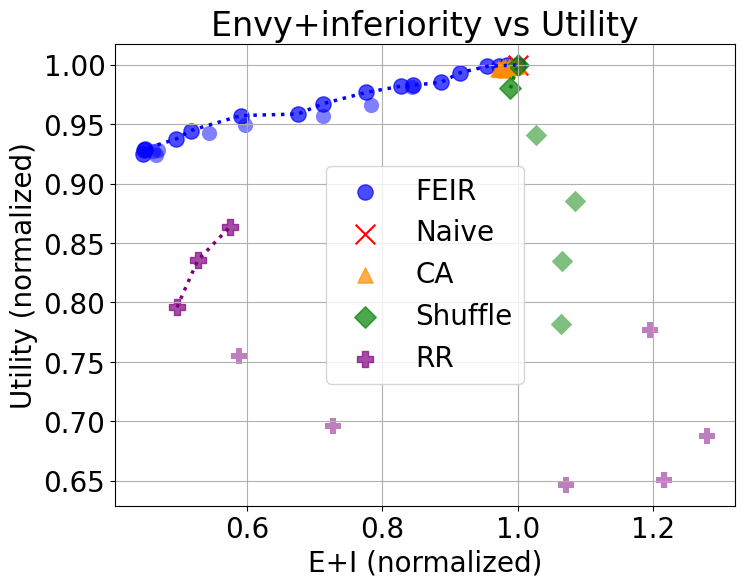}
    \caption{Synthetic $S \neq U$.}      
        \label{fig:ei_vs_u_su}
    \end{subfigure} 
    \hfill
    \begin{subfigure}[t]{0.24\textwidth}  
        \centering 
        \includegraphics[height=26mm]{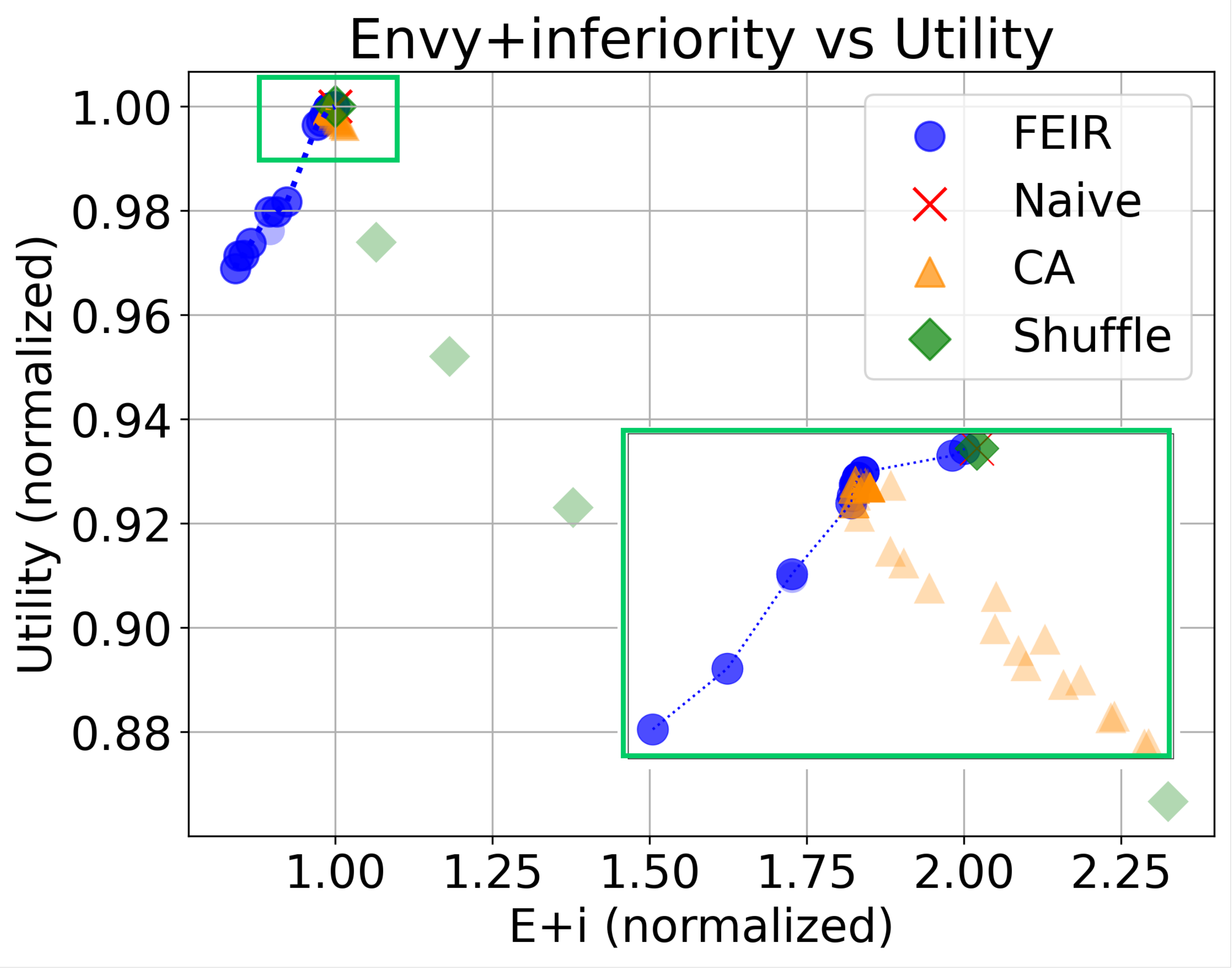}
  
    \caption{Synthetic $100 \times 20$.}      
        \label{fig:ei_vs_u_100by20}
    \end{subfigure} 
    \hfill
    \begin{subfigure}[t]{0.24\textwidth}
        \centering
        \includegraphics[height=26mm]{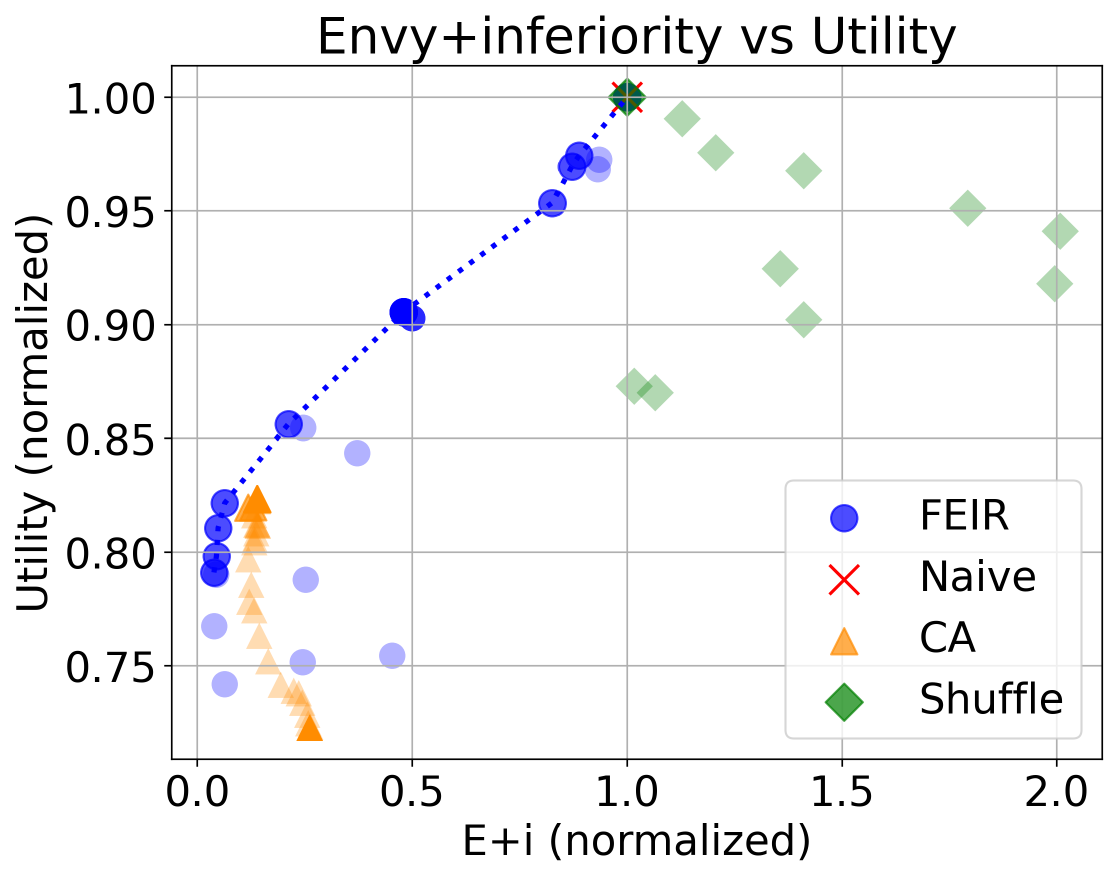}
   
    \caption{IG.}  
        \label{fig:ei_vs_u_skgroup}
    \end{subfigure}
    \hfill
    \begin{subfigure}[t]{0.24\textwidth}
        \centering
        \includegraphics[height=26mm]{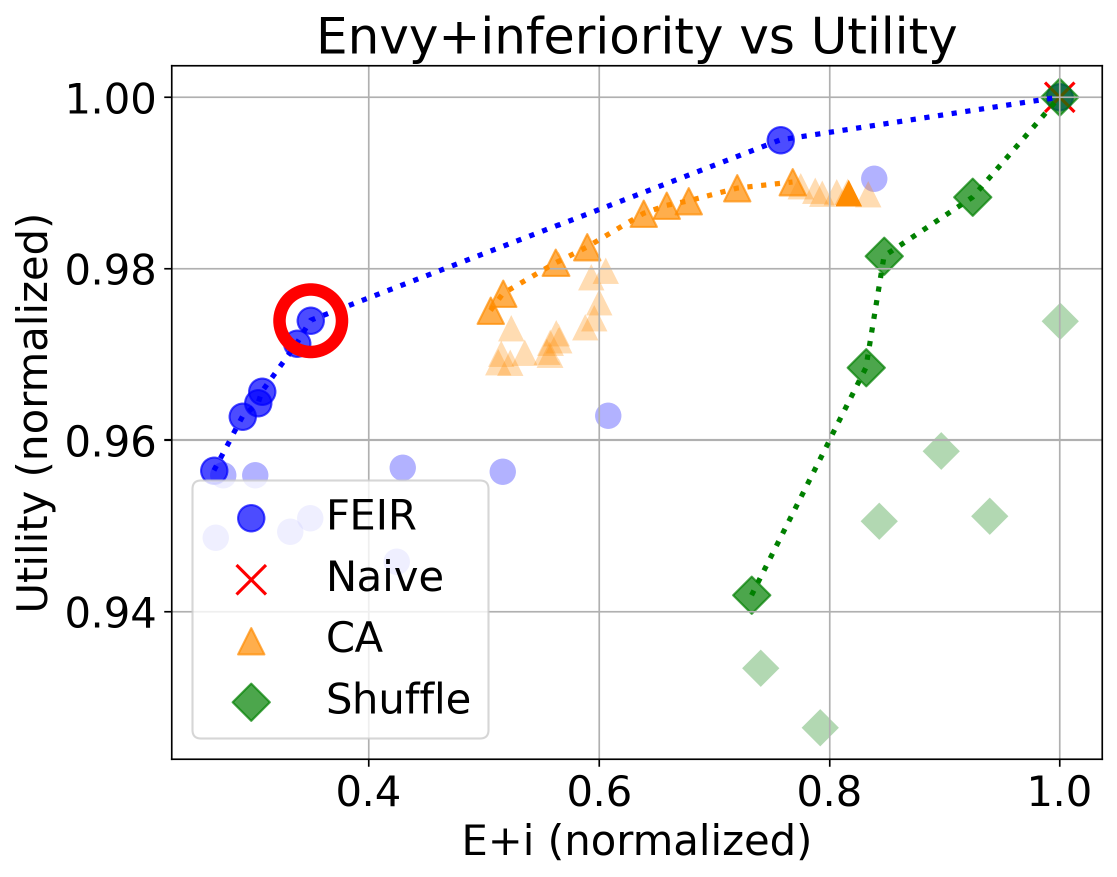}
  
    \caption{UG}  
        \label{fig:ei_vs_u_pplgroup}
    \end{subfigure}
\\
    \begin{subfigure}[t]{0.24\textwidth}
        \centering
        \includegraphics[height=26mm]{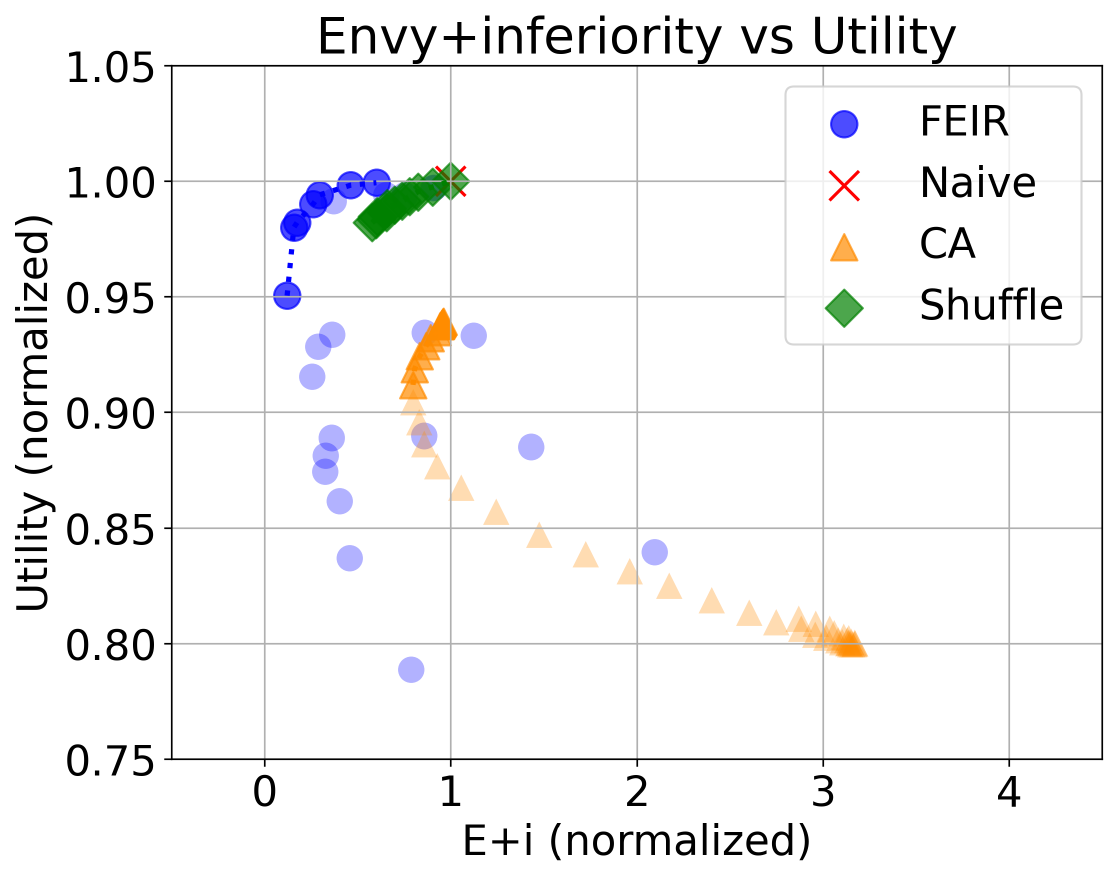}
  
    \caption{VDAB small $k=5$.}  
        \label{fig:ei_vs_u_vdabk5}
    \end{subfigure}
    \hfill
    \begin{subfigure}[t]{0.24\textwidth}
        \centering
        \includegraphics[height=26mm]{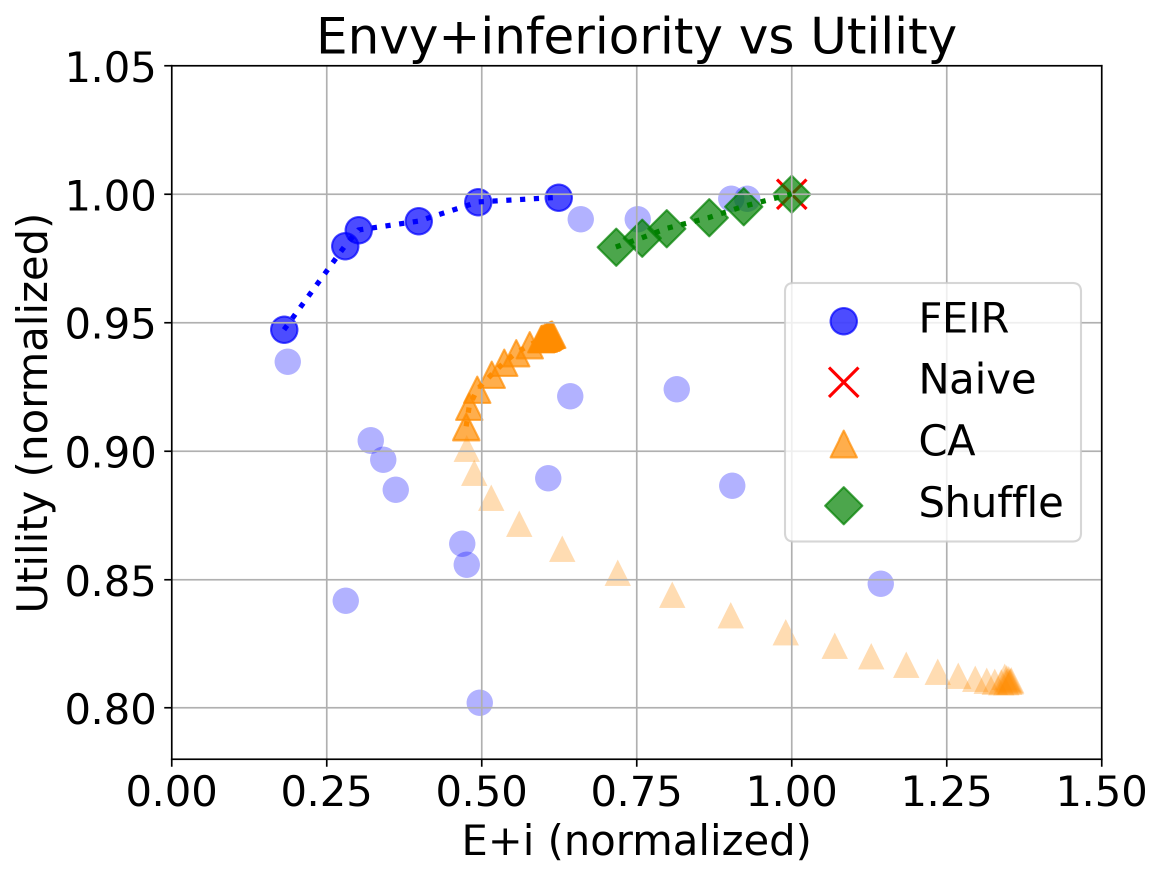}  
    \caption{VDAB small $k=50$.}  
        \label{fig:ei_vs_u_vdabk50}
    \end{subfigure}
    \hfill
    \begin{subfigure}[t]{0.24\textwidth}  
        \centering 
        \includegraphics[height=26mm]{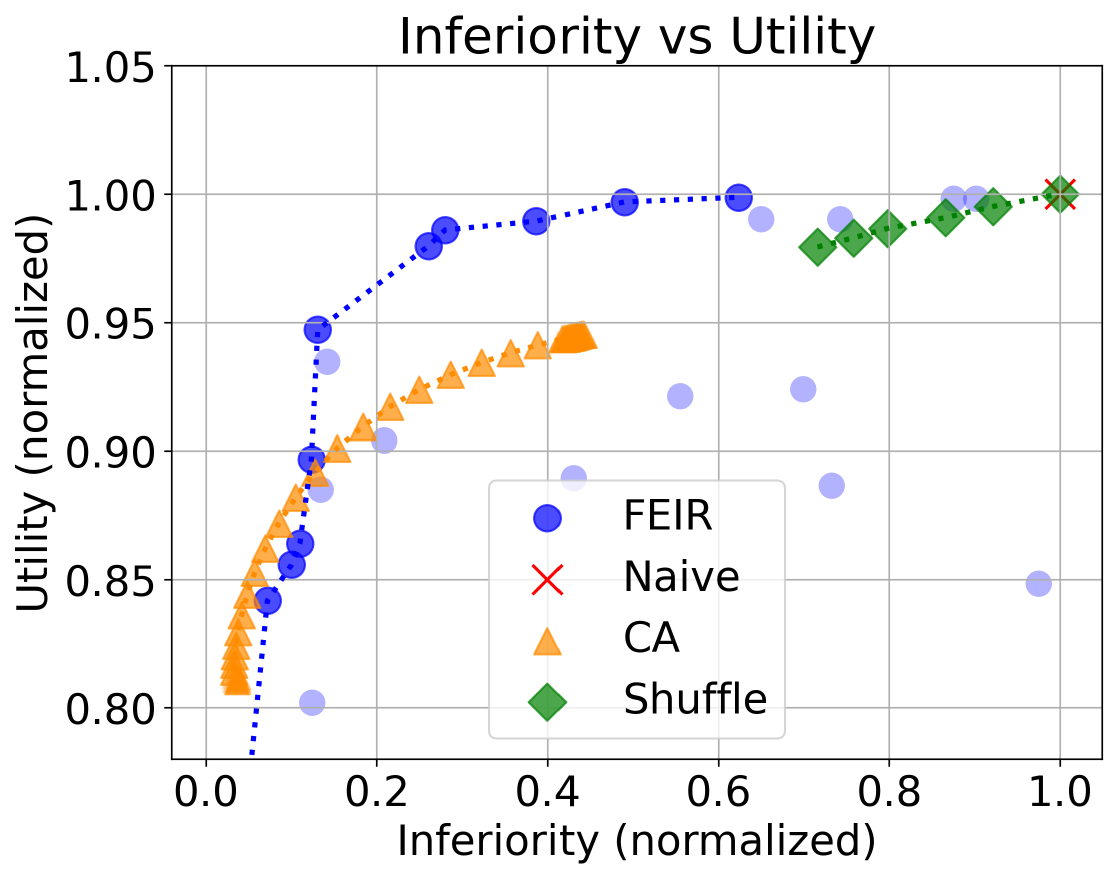}
    \caption{VDAB small: Inferiority vs utility $k=50$.}      
        \label{fig:i_vs_u_vdabk50}
    \end{subfigure}
    \hfill
    \begin{subfigure}[t]{0.24\textwidth}
        \centering
        \includegraphics[height=26mm]{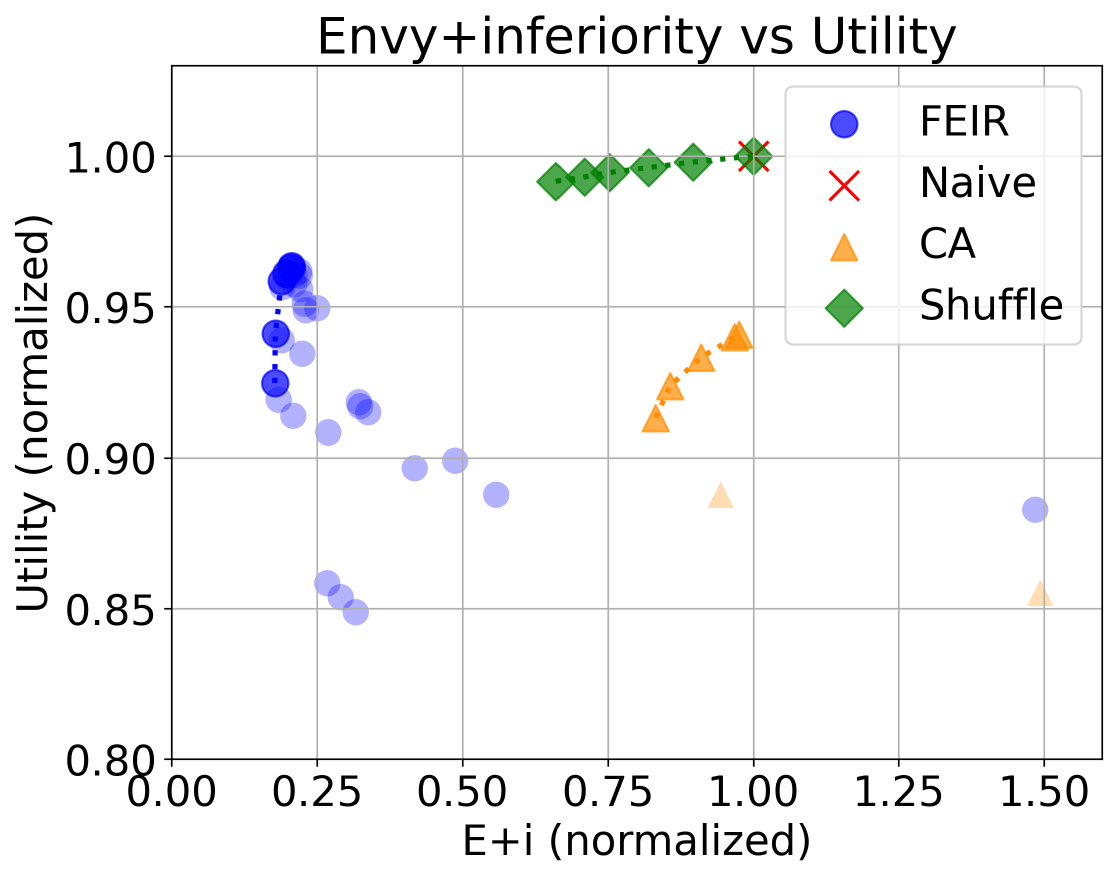}
    \caption{VDAB large $k=20$.}  
        \label{fig:ei_vs_u_vdabl}
    \end{subfigure}
\caption{Selected Pareto frontiers trading-off envy, inferiority and utility (upper-left better). (b): upper right region zoomed in. (d): The circled \texttt{FEIR} solution decreases inferiority of \textbf{both} user groups from Naive: the advantageous group \textbf{0.094 $\rightarrow$ 0.082} and the other \textbf{4.251 $\rightarrow$ 1.152}. %
(h): user-item sampling with a sample size about $\frac{1}{30}$ of the total users and $\frac{1}{70}$ items.\label{fig:eiu_synthetic}}
\end{figure*}

\subsection{Results}\label{sec:results}
\subsubsection{Fairness versus utility trade-offs (\ref{enum:rq1})}\label{sec:result_eiu_tradeoffs}
Our proposed method, \texttt{FEIR}, and the baseline methods were evaluated on synthetic and real-world datasets. The results indicate that both \texttt{FEIR} and CA can consistently improve fairness over the naive recommendation approach, while sacrificing some utility.
By varying the hyperparameters for the methods, different trade-offs between fairness and utility were achieved. To compare the results, we plotted each solution as a point on a graph with (un)fairness as the $x$-coordinate and utility as the $y$-coordinate, and drew the Pareto frontiers. 

\textbf{\textsc{Synthetic datasets.}}
\texttt{FEIR} is clearly the best (Fig.~\ref{fig:ei_vs_u_skgroup}, \ref{fig:ei_vs_u_pplgroup}), followed by CA, although the latter tends to cover a smaller solution region. RR scarifies too much utility for fairness (Fig.~\ref{fig:ei_vs_u_su}). Shuffle performs unstably.
 
Interestingly, a closer look at one of our solutions for \textbf{UG} shows that \texttt{FEIR} can simultaneously decrease the inferiority for both user groups (Fig.~\ref{fig:ei_vs_u_pplgroup}), which is desirable as it does not require sacrifices from one group to benefit the other.

When recommending items using the naive recommendation strategy with the random \emph{synthetic datasets with varying user-item ratios}, the inferiority increases with an increased ratio of users to items, indicating that the naive approach causes competitive disadvantages for users, and the more limitation the tenser the competition. 
CA does not decrease inferiority well when the number of items is not greater than the number of users; on the other hand, \texttt{FEIR} is able to find solutions with low inferiority  as seen in Fig.~\ref{fig:ei_vs_u_100by20}. When the number of items surpasses users, CA can also find solutions with low inferiority and high utility, but is still outperformed by \texttt{FEIR}
(corresponding figures included in our online supplementary.).

\textbf{\textsc{Real world datasets.}}
Data exploration confirms the existence of inferiority and competition caused by the naive recommendation. With increasing $k$s, the utility per recommendation decreases, and the inferiority and competition increase with a decelerating growth rate (see figures in our online supplementary).
The reason is that with a larger $k$, there are more overlapping recommendation and more competition, but also the average scores decrease with increasing $k$.

The \emph{VDAB small} and \emph{CareerBuilder} datasets show similar patterns in the relative performance of \texttt{FEIR} and CA. \texttt{FEIR} can decrease inferiority without reducing much utility or increasing envy, while CA decreases inferiority but also increases envy and reduces utility, especially when the number of recommendations is small. Shuffle prioritize utility, but cannot reduce much unfairness (Fig.~\ref{fig:ei_vs_u_vdabk5} and \ref{fig:ei_vs_u_vdabk50}). 

\begin{figure*}[t]
     \begin{subfigure}[t]{0.24\textwidth}
        \centering
        \includegraphics[height=26mm]{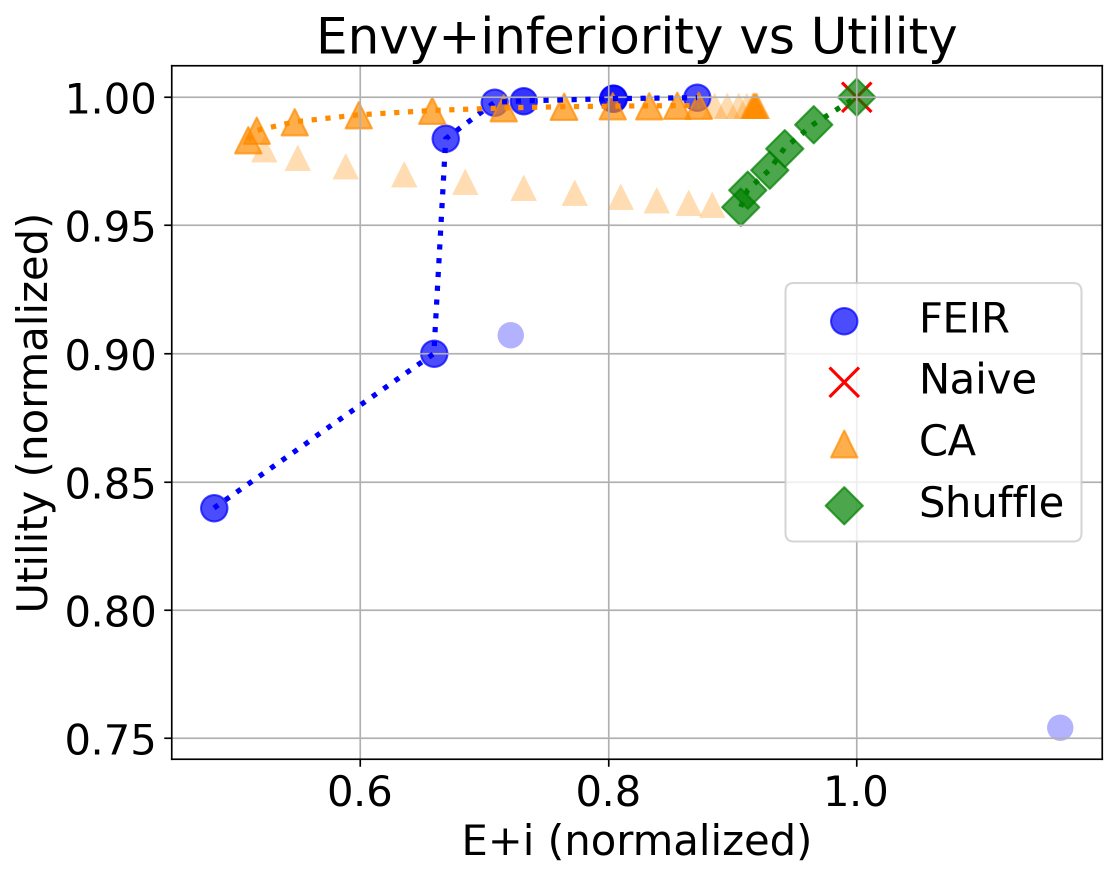}
  
    \caption{Zhilian all users $k=50$. }  
        \label{fig:ei_vs_u_zlk50}
    \end{subfigure}
    \hfill
    \begin{subfigure}[t]{0.24\textwidth}  
        \centering 
        \includegraphics[height=26mm]{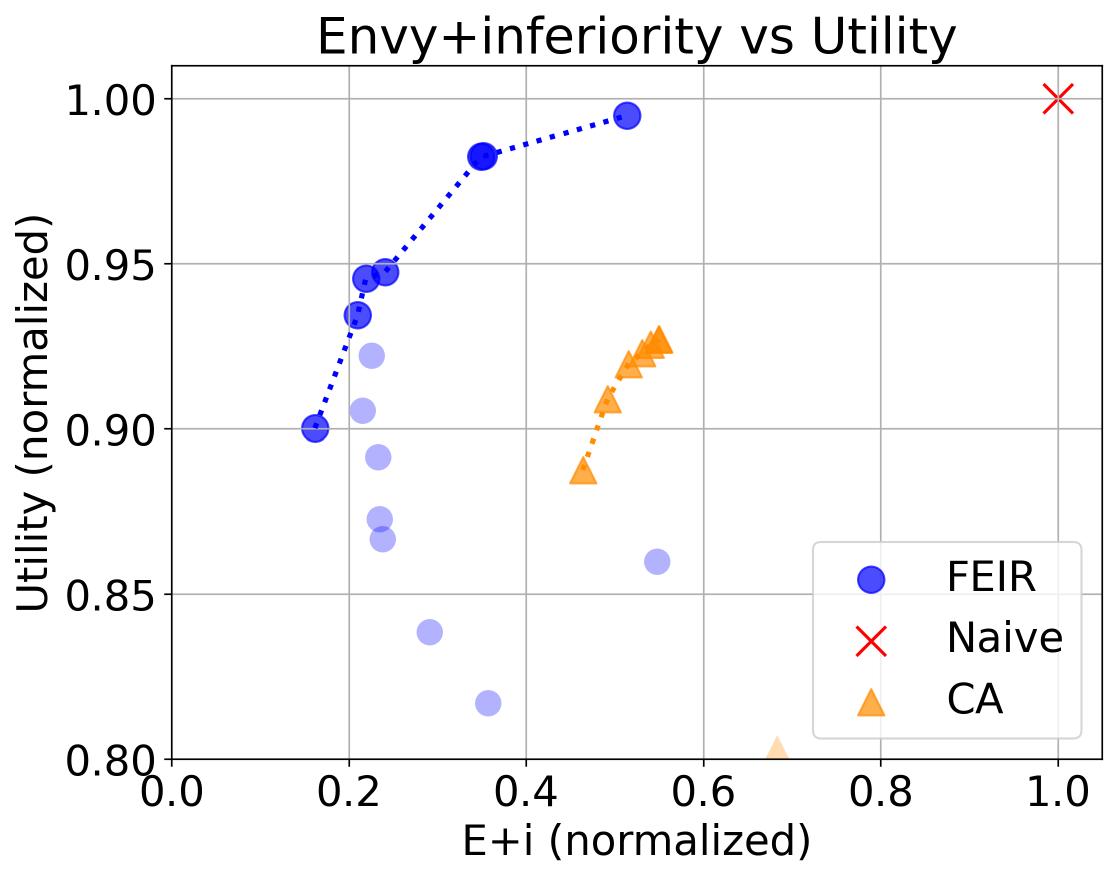}
    \caption{Zhilian one user cluser.}      
        \label{fig:ei_vs_u_zlc2k50}
    \end{subfigure}
    \hfill
    \begin{subfigure}[t]{0.24\textwidth}  
        \centering 
        \includegraphics[height=26mm]{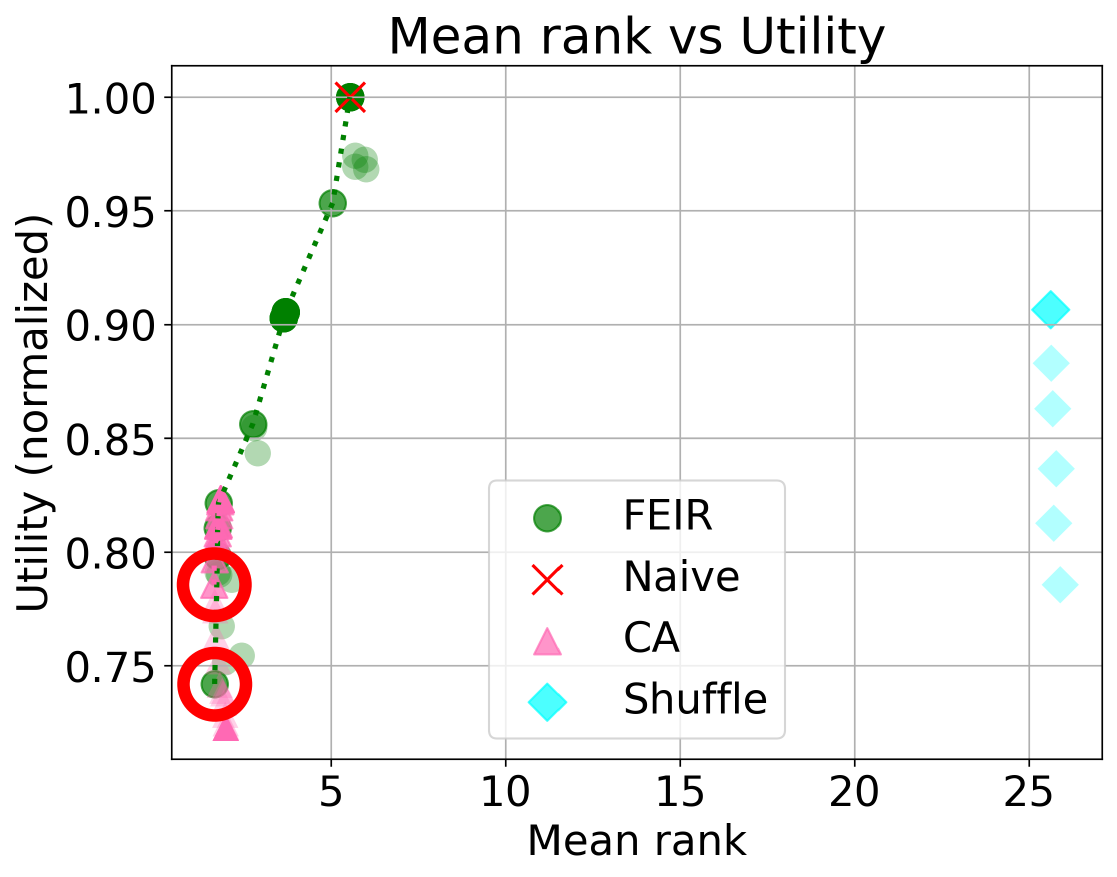}

    \caption{IG: rank.}      
        \label{fig:mr_vs_u_skgroup}
    \end{subfigure}    
    \hfill
    \begin{subfigure}[t]{0.24\textwidth}  
        \centering 
        \includegraphics[height=26mm]{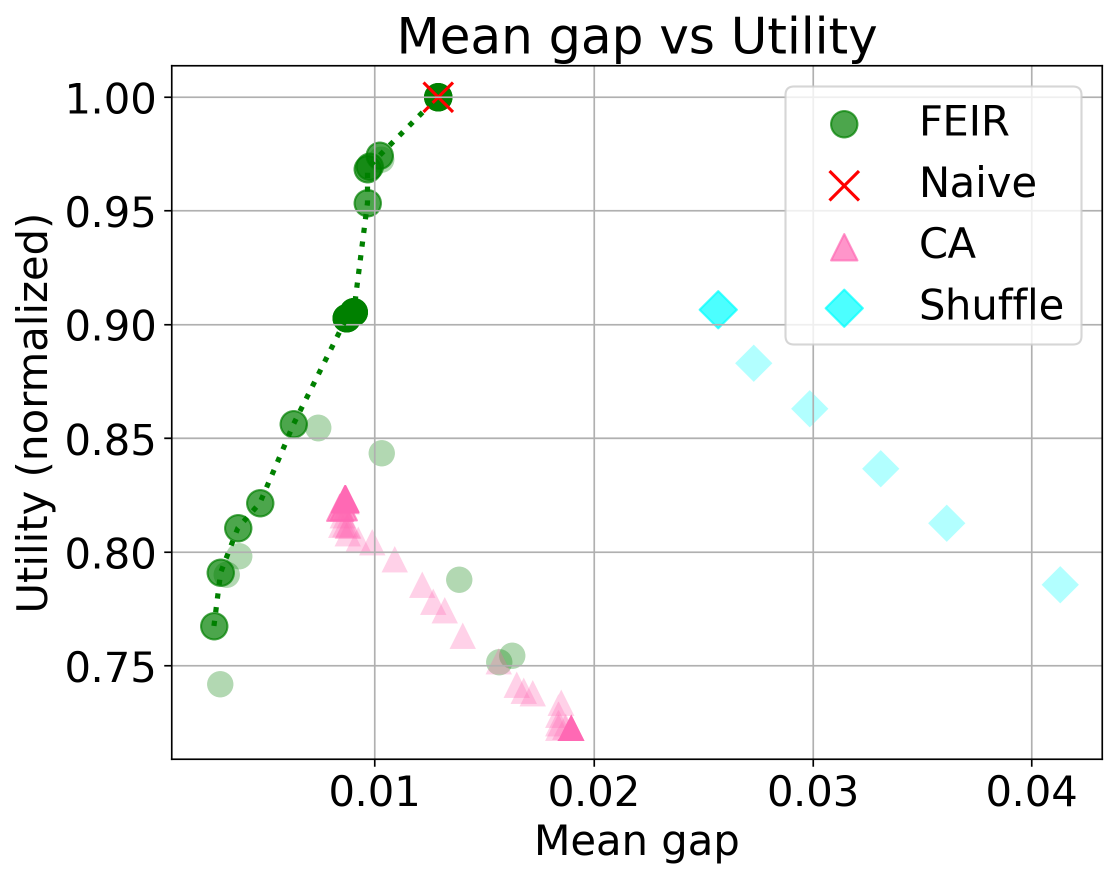}   
    \caption{IG: gap.}      
        \label{fig:ms_vs_u_skgroup}
    \end{subfigure} 
    \\
    \begin{subfigure}[t]{0.24\textwidth}  
        \centering 
        \includegraphics[height=26mm]{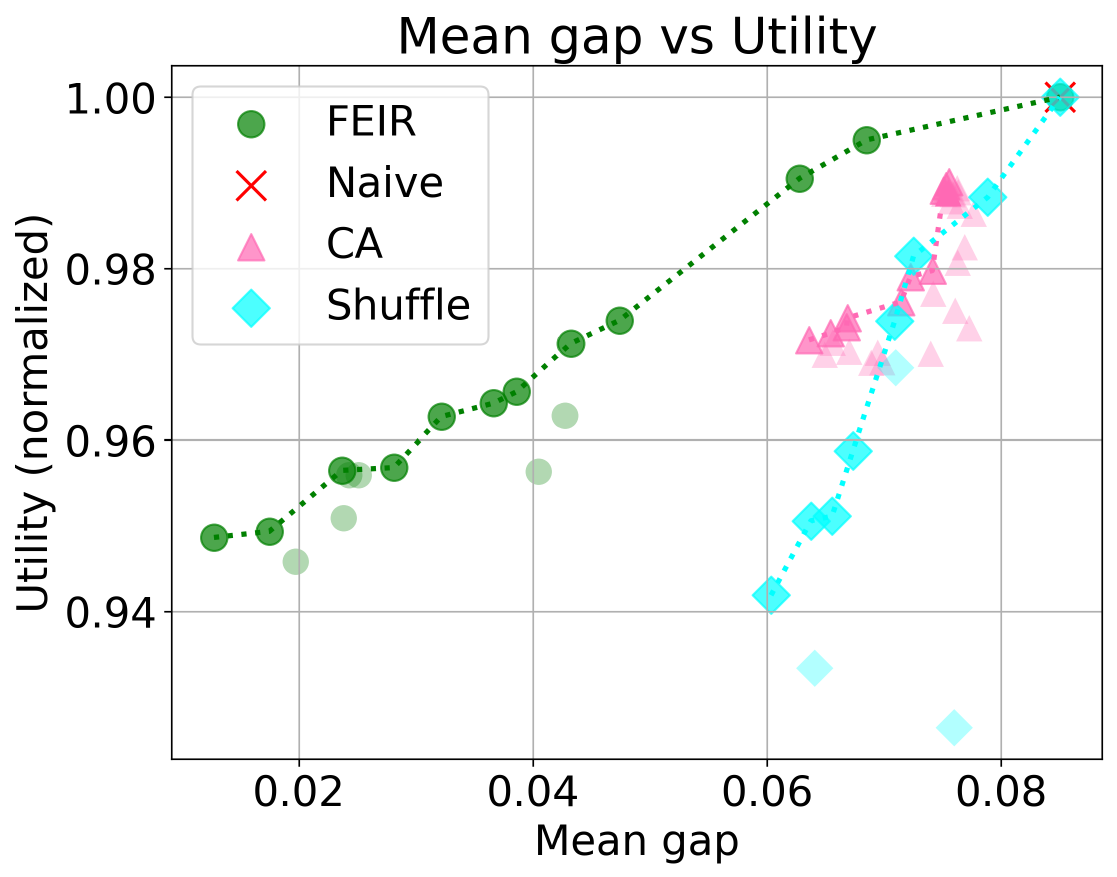}    
    \caption{UG.}      
        \label{fig:ms_vs_u_pplgroup}
    \end{subfigure}
    \hfill
    \begin{subfigure}[t]{0.24\textwidth}  
        \centering 
        \includegraphics[height=26mm]{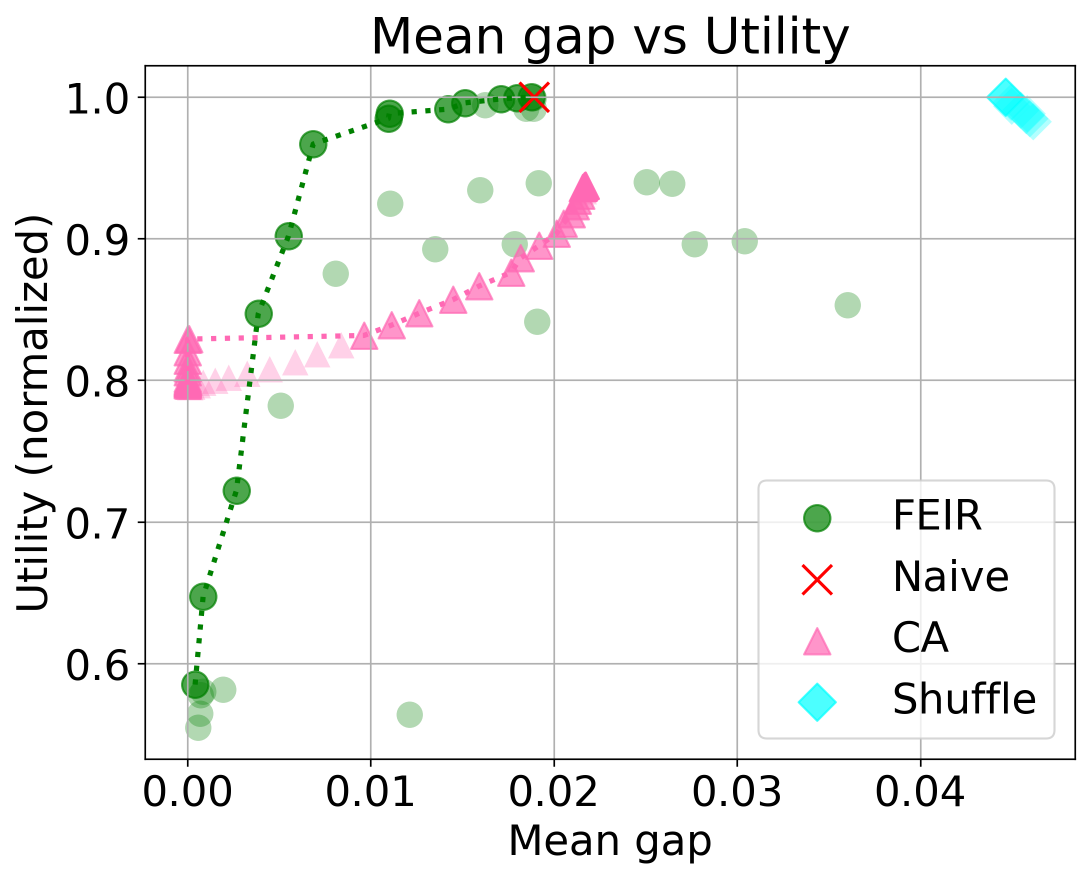}
   
    \caption{VDAB small $k=5$.}      
        \label{fig:ms_vs_u_vdabk5}
    \end{subfigure}  
    \hfill
    \begin{subfigure}[t]{0.24\textwidth}  
        \centering 
        \includegraphics[height=26mm]{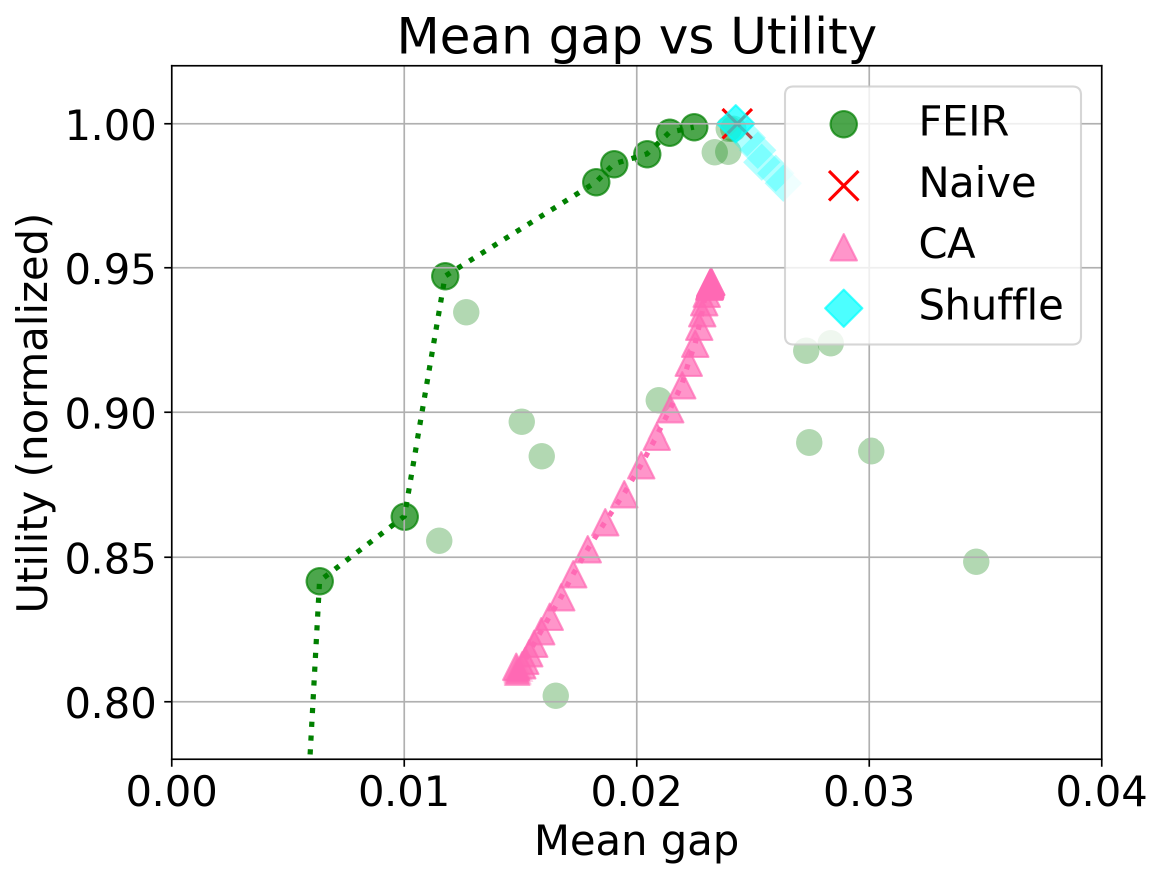}
   
    \caption{VDAB small $k=50$.}      
        \label{fig:ms_vs_u_vdabk50}
    \end{subfigure}     
    \hfill
    \begin{subfigure}[t]{0.24\textwidth}  
        \centering 
        \includegraphics[height=26mm]{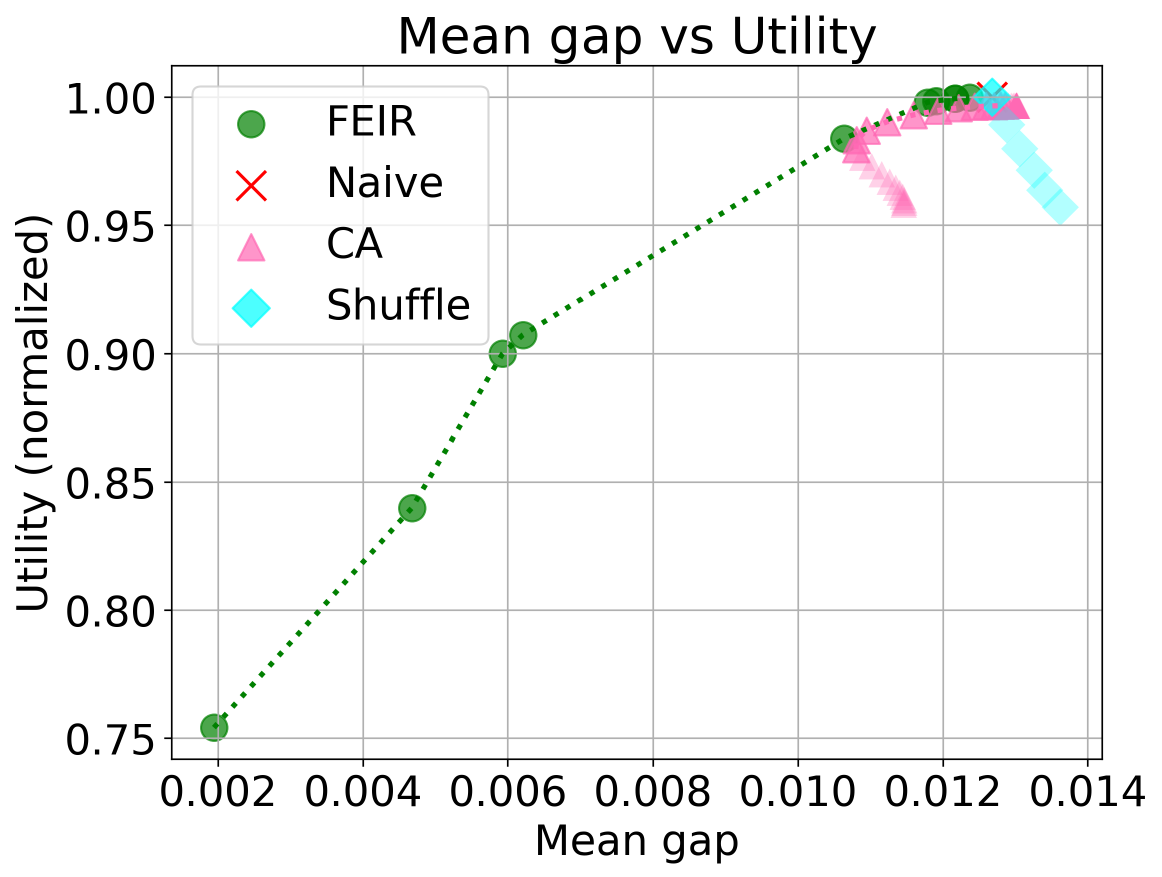}    
    \caption{Zhilian $k=50$.}      
        \label{fig:ms_vs_u_zlk50}
    \end{subfigure} 
\caption{Compare (a) with (b): \texttt{FEIR}'s performance on Zhilian dataset is not ideal when trained with \emph{user sampling}, but \texttt{FEIR} outperforms CA when trained without sampling. (c)-(h): Selected Pareto frontiers trading off competition and utility (upper-left better). (c): Two solutions with close mean ranks are circled. The mean suitability gap of \texttt{FEIR} is \textbf{0.003} while CA's is \textbf{0.012}.\label{fig:competition_vdab}}
\end{figure*}

With large $k=50$, \texttt{FEIR}'s performance in reducing unfairness is not as good as CA for the \emph{Zhilian} dataset when user sampling is used (Fig.~\ref{fig:ei_vs_u_zlk50}). Nonetheless, \texttt{FEIR} performs better than CA when trained on smaller subsets of users that can be processed in a single batch as seen in Fig.~\ref{fig:ei_vs_u_zlc2k50}. This suggests that the loss functions are effective, but the decreased performance is most likely due to the optimization process or some unique characteristics of the Zhilian dataset, which is left for future work.

\texttt{FEIR} performed well on the \emph{VDAB large} dataset, even with a sample size relatively small to the total numbers as show in Fig.~\ref{fig:ei_vs_u_vdabl}.

\begin{table}[t]
\centering
\caption{Comparison of the Pareto frontiers trading off fairness metrics with utility for the \textbf{CareerBuilder} dataset with varying $k$s.  The reference point for calculating the HVs is [1, 0.95]. The better results are marked bold.}
\label{table:cb_pareto_eiu}
\begin{tabular}[t]{c|cc|cc|cc|cc}
\toprule
\multirow{2}{*}{$k$} & \multicolumn{2}{c|}{HV($g$ vs $u$)} & \multicolumn{2}{c|}{HV($i$ vs $u$)} & \multicolumn{2}{c|}{$\min(g | 0.95)$} & \multicolumn{2}{c}{$\min(i | 0.95)$} \\ \cmidrule{2-9} 
                   & FEIR               & CA    & FEIR               & CA    & FEIR                & CA      & FEIR            & CA \\ \hline
1                  & \textbf{0.043}     & 0.013        & \textbf{0.048}     & 0.031      & \textbf{0.140}      & 0.642       & \textbf{0.006}  & \textbf{0.006}          \\
5                  & \textbf{0.042}     & 0.024        & \textbf{0.045}     & 0.031      & \textbf{0.138}      & 0.365       & \textbf{0.049}  & 0.127          \\
10                 & \textbf{0.041}     & 0.026      & \textbf{0.044}     & 0.032      & \textbf{0.143}      & 0.321       & \textbf{0.081}  & 0.142          \\
20                 & \textbf{0.039}     & 0.027      & \textbf{0.042}    & 0.031      & \textbf{0.185}     & 0.321       & \textbf{0.104}  & 0.175         \\
50                 & \textbf{0.034}     & 0.026     & \textbf{0.035}     & 0.029      & \textbf{0.278}      & 0.357       & \textbf{0.248}  & \textbf{0.248}          \\
100                & \textbf{0.029}     & 0.025      & \textbf{0.030}     & 0.027      & \textbf{0.367}      & 0.392       & 0.333           & \textbf{0.324}  \\
\bottomrule
\end{tabular}
\end{table}

Quantitative comparisons of the Pareto frontiers generated by \texttt{FEIR} and CA for the VDAB small, Zhilian and CareerBuilder datasets with various $k$ values support these observations. We only present the results for \emph{CareerBuilder} dataset here in Table~\ref{table:cb_pareto_eiu} due to space limitation.

\subsubsection{Competition faced by users (\ref{enum:rq2})}\label{sec:result_competition}
In general, CA is capable of achieving a low mean rank (Fig.~\ref{fig:mr_vs_u_skgroup}), but always a much higher mean gap compared to \texttt{FEIR} (Fig.~\ref{fig:ms_vs_u_skgroup}, \ref{fig:ms_vs_u_pplgroup}).
We argue that \texttt{FEIR} is more desirable. A recommendation with a low mean rank but a large mean suitability gap suggests that, although a user does not have many competitors, the competitors she does have are much better hence much more likely to defeat this user. For example, consider a job seeker $a_i$ with a suitability score of $0.7$ for a certain job. CA tends to recommend this jobs to only one other job seeker with a score of $0.99$, and on the other hand \texttt{FEIR} may recommend this job to three other job seekers with scoring $0.69$, $0.74$, $0.8$ respectively. It is reasonable to believe that \texttt{FEIR} gives user $a_i$ a better chance of getting hired, especially when considering that in reality, one would not apply for all recommended jobs. 
Shuffle performs almost always the worst.

When recommending a small number of jobs from a large pool, CA sometimes recommends non-overlapping jobs to each user, resulting in trivial solutions with no competition but decreased utility, as seen in the left most region of Fig.~\ref{fig:ms_vs_u_vdabk5}. However, \texttt{FEIR} can provide solutions with higher utility. As $k$ increases, it becomes harder to give non-overlapping recommendations for CA such that \texttt{FEIR} always gives a lower suitability gap (Fig.~\ref{fig:ms_vs_u_vdabk50}, \ref{fig:ms_vs_u_zlk50}).

\begin{table}[t]
\centering
\setlength{\tabcolsep}{1pt}
\caption{Comparison of the Pareto frontiers trading off competition metrics with utility for the \textbf{VDAB small} data with varying $k$s. The reference point for calculating the HV(rank vs u) being [50, 0.9] means the reference value of the mean rank is 50 and the normalized utility 0.9, and for HV(gap vs u) [0.03, 0.9] means the reference value of the mean suitability gap is 0.03.}
\label{table:vdab_pareto_competition}
\begin{tabular}{c|cc|cc|cc|cc}
\toprule
\multirow{2}{*}{k} & \multicolumn{2}{c|}{HV(rank vs u)} & \multicolumn{2}{c|}{HV(gap vs u)} & \multicolumn{2}{c|}{$\min(\text{rank}| 0.9)$} & \multicolumn{2}{c}{$\min(\text{gap} | 0.9)$} \\ \cmidrule{2-9} 
                   & FEIR                 & CA     & FEIR                & CA      & FEIR             & CA           & FEIR                 & CA      \\ 
\midrule
1                  & \textbf{4.747}       & 1.496       & \textbf{0.003}      & 0.0         & \textbf{1.286}   & 3.66              & \textbf{0.002}       & 0.017       \\
5                  & \textbf{4.415}       & 1.54        & \textbf{0.002}      & 0.0         & \textbf{2.731}   & 5.318             & \textbf{0.006}       & 0.02        \\
10                 & \textbf{4.103}       & 1.535       & \textbf{0.002}      & 0.0         & \textbf{4.599}   & 6.761             & \textbf{0.007}       & 0.021       \\
20                 & \textbf{3.592}       & 1.523       & \textbf{0.002}      & 0.0         & \textbf{8.317}   & 8.805             & \textbf{0.007}       & 0.022       \\
50                 & \textbf{2.463}       & 1.438       & \textbf{0.001}      & 0.0         & 15.462           & \textbf{11.28}    & \textbf{0.012}       & 0.021       \\
100                & \textbf{1.482}       & 1.291       & \textbf{0.001}      & 0.0         & 29.442           & \textbf{16.099}   & \textbf{0.018}       & 0.022       \\ 
\bottomrule
\end{tabular}
\end{table}

A quantitative comparison of the Pareto frontiers generated by \texttt{FEIR} and CA for the \emph{VDAB small} dataset with various $k$ values shows that \texttt{FEIR} is better than CA almost across the board, except for $\min(\text{rank}|0.9)$ with $k=50$ and $100$ (Table~\ref{table:vdab_pareto_competition}). The \emph{CareerBuilder} dataset has similar results with VDAB small where \texttt{FEIR} is better than CA in general, while \texttt{FEIR} shows less advantage over CA for \emph{Zhilian} (plots and tables in our online supplementary), as discussed in Section~\ref{sec:result_eiu_tradeoffs}.

\subsubsection{Item-side fairness}\label{sec:item_fairness}
\texttt{FEIR} improves the fairness to the items as the Gini index decreased greatly for all datasets after \texttt{FEIR} post-processing (Table~\ref{table:gini}).

Our \emph{code and supplementary} materials for more details and extra plots are publicly available at \url{https://github.com/aida-ugent/FEIR}.

\section{Related work}\label{sec:related_work}

This paper extends the growing literature on fairness in machine learning (e.g. \cite{dwork2012fairness,hardt2016equality,zafar2017parity,kusner2017counterfactual,yao2017parity,steck2018calibrated,wang2021user,abdollahpouri2020multistakeholder,do2021twosided,patro2020fairrec}).
Here we summarize the most directly related research.

\emph{Fairness when recommending items with limited availability.}
Particularly in the context of job recommendations, this is an increasingly active research area. Yet, the current literature mainly focuses on \emph{group level} disparity notions. For example, \citet{546/geyik2019fairness} proposed four deterministic reranking algorithms to mitigate biased prediction towards any sensitive job seeker group, and \citet{531/islam2021debias}  addressed gender bias in job recommendations by proposing a neural fair collaborative filtering model. In contrast to this existing work, we focus on fairness from the perspective of \emph{individual} users, rather than group level fairness. Other orthogonal research includes fairness for jobs and interdisciplinary studies (see recent survey by \citet{mashayekhi2022achallenge}). 

\emph{Competition and congestion in recommendation.} To the best of our knowledge, there has been no research at all on the concept of inferiority. Yet, \citet{naya2021designing} did study the related notion of \emph{congestion}, in the context of labor market. They proposed a congestion alleviation method, which reduces the intersection between the sets of jobs recommended to different job seekers. Congestion does not consider suitability (i.e. competitiveness) of users for their recommended jobs like inferiority does.

\emph{Envy-freeness in recommendation.} Inspired by the literature on social choice theory and fair resource allocation (e.g., \cite{foley1967resource,moulin2003fair,varian1974equity}), a few researchers recently introduced the notion of envy-freeness into the context of recommendation systems. \citet{do2022online} gave a generic individual-level definition of envy-freeness and cast the problem of auditing for such envy-freeness as an exploration problem in multi-armed bandits. Their focus is online \emph{evaluation} (auditing) of existing systems, while we aim to also \emph{minimize} envy as well as inferiority, using a post-processing method.  
\citet{patro2020fairrec} designed a modified Round-Robin algorithm to ensure fairness on the item side while guaranteeing envy-freeness up to one good (EF1) fairness for every user, and \citet{wu2021tfrom} extended this approach to producer fairness. Besides the fact that we do not share their focus on item-side fairness, their problem settings do not apply to limited resource recommendation because the users in their setting do not compete with each other. 

\begin{table}[t]
\centering
\caption{\texttt{FEIR} also improves the item-side fairness as the Gini indices of item exposure for all datasets are decreased after \texttt{FEIR} post-processing.}
\label{table:gini}
\begin{center}
\begin{tabular}[t]{ccccccc}
\toprule
Dataset                 & IG & UG & V(S) & V(L) & ZL & CB \\
\midrule
Gini index $\downarrow$  \% & 73 & 60 & 34 & 53 & 60 & 37 \\
\bottomrule
\end{tabular}
\end{center}
\end{table}

\section{Discussion and Conclusion}\label{sec:conclusion} 
Recommending items with limited availability to users has its own challenges and brings new fairness requirements not addressed in the existing literature. In this paper we proposed envy and inferiority as important fairness notions to fill the gap and presented a post-processing approach \texttt{FEIR} to improve the fairness of such recommendation settings. 

Our experiments on synthetic and real job recommendation datasets demonstrated that \texttt{FEIR} improves fairness by reducing the potential competitive disadvantage of users without significantly sacrificing utility. Importantly, our method \texttt{FEIR} is not limited to the labor market, but also promising in reducing user inferiority and competitive disadvantages in other real-world scenarios such as online dating, paper bidding systems, and education resources recommendation. 

Our work has limitations but also opens up new research opportunities. 
The actual competition and chances of getting any item depend on many factors beyond any recommendation system and hence beyond our scope. 
Also, emphasizing envy and inferiority does not make other existing fairness concerns any less important, nor the case that they can cover all new fairness requirements from the unique features of recommending limited resources. Rather, our findings create new opportunities for research to explore the relations among different fairness notions and identify other ignored dimensions of fairness in these settings.

Some alternative formulations of utility, envy and inferiority are possible. For example, disallowing repeated recommendation for a user, which involves further complexity in the probabilistic setting. It is also possible to modify the quantification of inferiority by taking the utility into account. The analysis and comparison of the current formulation and the alternatives would be interesting for future work.
Besides, the interests of recruiters could be further considered by adapting the optimization objective to include some metrics representing the suitability of candidates. The dynamics between job seeker side and recruiter side fairness is another future direction worth exploring.  

\begin{acknowledgments}
The research leading to these results has received funding from the European Research Council under the European Union's Seventh Framework Programme (FP7/2007-2013) (ERC Grant Agreement no. 615517), and under the European Union’s Horizon 2020 research and innovation programme (ERC Grant Agreement no. 963924), from the Special Research Fund (BOF) of Ghent University (BOF20/IBF/117), from the Flemish Government under the ``Onderzoeksprogramma Artificiële Intelligentie (AI) Vlaanderen'' programme, and from the FWO (project no. G0F9816N, 3G042220). Part of the experiments were conducted on pseudonimized HR data generously provided by VDAB (Vlaamse Dienst voor Arbeidsbemiddeling en Beroepsopleiding).
\end{acknowledgments}

\bibliography{ref}

\end{document}